\documentclass[aps,prl,superscriptaddress,onecolumn,notitlepage]{revtex4-2}

\usepackage{amsmath,amssymb,graphicx,color,bm,ulem}

\begin{document}

\title{Light-induced injection of hot carriers from gold nanoparticles to carbon wire bundles}

\author{Stella Kutrovskaya}
\affiliation{Abrikosov Center for Theoretical Physics, Moscow Institute of Physics and Technology (National Research University), 141701 Institutskii per., Dolgoprudnyi, Moscow Region, Russia}
\affiliation{Department of Physics and Applied Mathematics, Stoletov Vladimir State University, 600000 Gorkii street, Vladimir, Russia}

 \author{Igor Chestnov}
 \affiliation{ITMO University, St. Petersburg 197101, Russia}
\affiliation{Department of Physics and Applied Mathematics, Stoletov Vladimir State University, 600000 Gorkii street, Vladimir, Russia}

\author{Anton Osipov}
\affiliation{Department of Physics and Applied Mathematics, Stoletov Vladimir State University, 600000 Gorkii street, Vladimir, Russia}

\author{Vlad Samyshkin}
\affiliation{Department of Physics and Applied Mathematics, Stoletov Vladimir State University, 600000 Gorkii street, Vladimir, Russia}

\author{Anastasya Lelekova}
\affiliation{Department of Physics and Applied Mathematics, Stoletov Vladimir State University, 600000 Gorkii street, Vladimir, Russia}

\author{Alexey Povolotskiy} 
\affiliation{Institute of Chemistry, St. Petersburg State University, 198504, Ulyanovskaya str. 5, St. Petersburg, Russia}

\author{Xiaoqing Zhou}
\affiliation{School of Science, Westlake University, 18 Shilongshan Road, Hangzhou 310024, Zhejiang Province, China}
\affiliation{Institute of Natural Sciences, Westlake Institute for Advanced Study, 18 Shilongshan Road, Hangzhou 310024, Zhejiang Province, China}

\author{Alexey Kavokin}
 \affiliation{School of Science, Westlake University, 18 Shilongshan Road, Hangzhou 310024, Zhejiang Province, China}
\affiliation{Institute of Natural Sciences, Westlake Institute for Advanced Study, 18 Shilongshan Road, Hangzhou 310024, Zhejiang Province, China}
\affiliation{Spin Optics Laboratory, St. Petersburg State University, 198504, Ulianovskaya street 1, St. Petersburg, Russia}

\author{ Alexey Kucherik}
\affiliation{Department of Physics and Applied Mathematics, Stoletov Vladimir State University, 600000 Gorkii street, Vladimir, Russia}

\begin{abstract} We observed a light-induced enhancement of the tunneling current propagating through an array of parallel carbon chains anchored between gold nanoparticles (NPs). In the presence of laser radiation characterized by a wavelength close to the plasmon resonance of the NPs, the current-voltage characteristics of carbon bundle tunnelling junctions demonstrate a pronounced asymmetry between positive and negative bias values. Such an asymmetry is typical for a Schottky junction, in general. The resistance of the tunnel junction decreases with the increase of the optical pumping intensity. We associate the observed effect with an injection of `hot' carriers created in Au NPs due to the decay of the surface plasmons accompanied by the charge transfer to the carbon bundles. The observed phenomenon can be used for non-resonant excitation of excitonic states in low-dimensional carbon-based structures for single-photon emission, as well as for photovoltaic applications.
\end{abstract}

\maketitle

\section*{Introduction}

\noindent  Molecular electronics allows creating nanoscale devices on-a-chip that would be infeasible within the conventional silicon-based platform \cite{Mathew}. The spatial extent of molecular structures suitable for electronic applications ranges from the subnanometer scale upto the scale of hundreds of nanometers. In this way, molecular electronics is bridging the gap between lithographic and atomic structures. Among various potential applications of molecular devices, the use of long linear atomic-scale structures as ultra-thin electrical wires is of a particular interest. The nanoscale charge transport in linear molecular systems consisting of atomic chains placed between bulk electrodes is in the focus of this study. 

The control of interfaces between long linear atom-scale chains and the surfaces of metallic electrodes is crucial for the integration of carbon molecules into modern nanoelectronic devices \cite{Compagnini}. The interaction of carbon filaments with metal surfaces  has been studied in Refs.~\cite{Kaun, Li, Wold, Land} which paves the way to a number of nanoelectronic applications. 
Besides their applications as nanoscale conducting wires, conjugated oligomers are highly promising for the realisation of nano-switches as they demonstrate a negative differential resistance even at the single-molecule level \cite{Crljen}.

The variety of proposed nanowire designs grows very rapidly nowadays \cite{Machin}. Among the most fertile material platforms for nanowire applications are transition metal oxides, conductive polymers, carbon etc. Carbon, in particular, is known for providing a prototypical example of 1D-materials -- carbon nanotubes. Besides that, straight linear chains of carbon attract a lot of attention at present due to their unique electrophysical and optical properties \cite{ Cataldo, Tabata, Eisler, Compagnini2005, DUrso}. To cite an example, the new effect of strong delocalization of $\pi$–electrons in a polyine chain on the quantum transport of electrons in such structures has been reported in \cite{Zanolli,Maa}. The electronic properties of such linear structures crucially depend on the inter-atom spacing which can be tuned by applying a mechanical strain \cite{Cretu2013, Deretzis}. The strain-assisted switching of conductivity of the linear carbon filament was demonstrated in \cite{LaTorre2015}. The theoretical analyses \cite{Artyukhov} of this phenomenon highlights the impact of the bond length alternation on the band gap energy in a linear carbon chain.

The experimental study of electronic properties of single carbon chains is challenging for a simple reason: the carbon chains are never really single. Typically, they form bundles attached to metal NPs incorporated in a studied electric circuit. In order to reveal the quantum confinement effects in nano-wires one should look for nano-size structures, but conventional electric current measurements are extremely challenging on such a short scale. Still, many publications have been devoted to the synthesis of linear structures based on oligomers, polyyne filaments enclosed in carbon nanotubes or flakes consisting of substantially dissimilar carbon phases. Generally, adding endcapped  groups leads to the strong changes of the eigenstates of oligomers. Moreover, carbon filaments show the effect of negative differential resistance (NDR) in the heterostructures composed by metal contacts and carbon nanotubes \cite{Kaun, Khoo}. Interestingly the measured resistance values appeared to be significantly below the expected theoretical limit \cite {Yuzvinsky}. The transport properties of arrays of parallel polyyne chains and their dependence on the distance between the chains and the contact with the massive electrode have been investigated in \cite{Wang}.

In the present work, we study the tunneling transport in a multicompound thin film of carbon threads formed by polyyne oligomers and confined between gold nanoparticles. The current-voltage characteristics (CVCs) obtained by the scanning tunneling microscopy (STM) demonstrate an asymmetric behavior that would be characteristic for a semiconductor. In contrast, the nanoparticles show typical metallic CVCs. The irradiation of the film at the frequency close to the plasmon resonance of gold nanoparticles creates `hot' carriers. The value of the tunnelling current collected from long linear carbon chains (LLCCs) away from Au NPs depends on the excitation intensity which reflects a correlation of the intensity of a local plasmon field with the density of photogenerated carriers. The possible application of the observed effects for the realization of a field-effect transistor with a light-induced Schottky gate is discussed here.

\section{Results}

\subsection*{The linear carbon chains: formation and study of thin films with embedded gold NPs}

\begin{figure}
\includegraphics[width=0.75\linewidth]{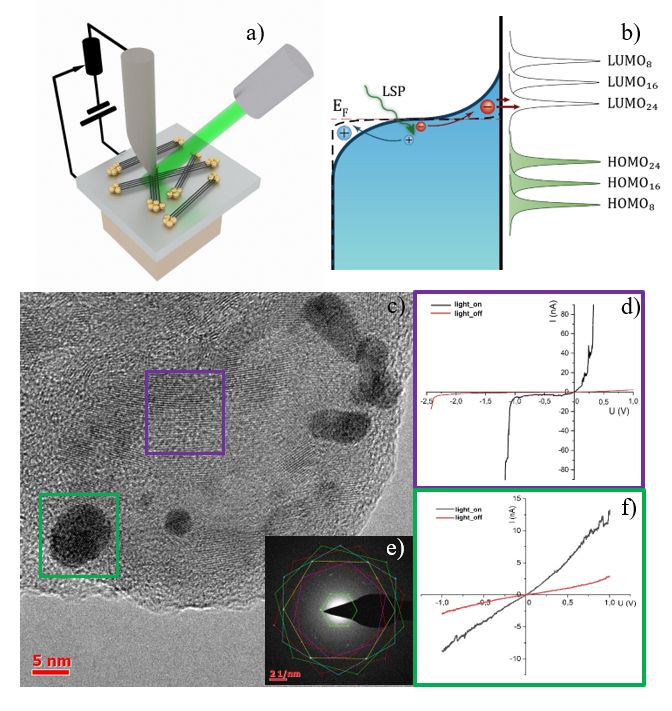}
\caption{The characterization of thin C-Au films: (a) a sketch of an experimental setup for measuring the current-voltage characteristics of a tunnel junction: the air gap between a thin film and a conductive probe are illuminated by 532 nm laser. Panel (b) depicts a schematics of the energy level structure of the metal NP-carbon wire junction illustrating plasmon-assisted generation of hot carriers and their injection into the wire. (c) TEM-image of the deposited layer showing orientation of the carbon filaments and metal nanoparticles (black spheres). The framed fragments correspond to the regions where the $I$--$U$ characteristics shown in panels (d) and (f) have been measured. In particular, the data shown on the purple panel (d) have been collected from the region containing the carbon chains only. The tunnel current collected from the gold NP is shown in the green panel (f). (e) shows an electron diffraction pattern with six embedded hexagons encircled by colour lines illustrating the multilayer thin film structure.}\label{fig1} 
\end{figure}

A colloidal system consisting of a mixture of gold NPs and LLCCs was manufactured using the method of laser ablation in a liquid \cite{Samyshkin}. The colloidal solution was deposited on a substrate by sputtering in the static electric field \cite{Kutrovskaya}. It yields a thin film of randomly oriented elongated bundles of parallel carbon chains anchored to the nanonaprticles at both ends as it is schematically shown in Fig.~\ref{fig1}a. This bundle configuration obeys energy minimization in the system of polyyne filaments, where the Van der Waals interaction between adjacent filaments stabilizes the whole structure compensating the overall distortion \cite{Mandal}. For a direct visualization of the carbon wires end-capped with Au NPs, we used a high-resolution transmission electronic microscopy (HR-TEM) technique, see Fig.~\ref{fig1}c. The analysis of the deposited film using the electron diffraction confirms the formation of a crystal $sp$-hybridized phase characterised by the lattice constant $a=9.56$~{\AA} \cite{Kutrovskaya, Kutrovskaya20} (the central hexagon in Fig.~\ref{fig1}e). The auxiliary nested and rotated about the same center hexagons correspond to the polycrystalline structure characterised by a limited number of selected directions along which the deposited threads are oriented. It leads to a shift in the secondary diffraction maxima in the diffraction pattern. 

\subsection*{Electrophysical properties of thin films}

We study the conductivity properties of the deposited structures by measuring the tunneling current with use of the STM. The basic STM setup has been modified in order to gain a possibility of a precise  \textit{in-situ} excitation of the sample with the green laser radiation with $\lambda=532$~nm. First, we study films without laser excitation, see the red curves in Figs.~\ref{fig1}d,f. In this case, carbon bundles and metal nanoparticles demonstrate a similar behaviour: their $I$--$U$ curves are symmetric with respect to the bias inversion as it is expected for the metal-insulator-metal junction \cite{ChenBook}. Besides, both CVCs show almost linear $I(U)$-dependence as the tip-voltage is swept between $\pm 1$~V which is consistent with the wide band gap of carbyne wires \cite{Mostaani}. However, as the light is switched on, the I-V characteristics of the carbon wire films become asymmetric and they start resembling those of a semiconductor-metal junction, i.e. of the Schottky barrier. Besides, the sharp jumps of the conductivity appear at both positive and negative biases, see Fig.~\ref{fig1}d.  The current collected from the NP surface is also changed under light illumination.  However, in contrast to the previous case, it merely amplifies its magnitude while remaining symmetric with respect to the bias inversion, see Fig.~\ref{fig1}f.

The tunneling current demonstrates a strong dependence on the laser intensity. Figure~\ref{fig2} shows that the magnitude of the current flowing through the tip-bundle junction grows with the increase of the laser power at the fixed tip-to-sample distance. This behaviour is also revealed in the increase of the differential conductance $dI/dU$ measured at the positive bias, see the upper inset in Fig.~\ref{fig2}.

\section{Discussion}

The observed light-assisted tunnelling enhancement can be connected with excitation of an extra amount of free carriers. The tunneling current $I$ collected by the STM tip is described within the Landauer-B{\"u}ttiker approach for coherent transport \cite{ChenBook}:
\begin{equation}\label{Eq_LBcurrent}
I(U)=\frac{2e}{\hbar}\int^{\infty}_{-\infty}\left[f_{\rm tip}(E) – f_{\rm s}(E+eU) \right] \rho_{\rm tip}(E) \rho_{\rm s}(E+eU) |M(E)|^2 dE,
\end{equation}
where $f_{\rm tip}(E)$ and $f_{\rm s}(E)$ are the carrier distribution functions of the tip and the investigated surface (either LLCC bundle or NP), respectively; 
$\rho_{\rm tip}(E)$ and $\rho_{\rm s}(E)$ are respective local densities of states (LDOS), $M(E)$ is the tunneling matrix element governed by the tip-to-surface distance and $U$ is the tip voltage. Since the laser fields does not interact with the charges in the STM probe, the tip electrons remain at equilibrium both in the dark and under the light illumination, 
$f_{\rm tip}(E) = f_{\rm eq}(E)=\left[1+\exp\left((E-E_{\rm F})/k_{\rm B} T\right)\right]^{-1}$,
 where $E_{\rm F}$ if the Fermi level of the tip, $k_{\rm B}$ is the Boltzmann constant, $T$ is the temperature. 
 Therefore, the tunneling enhancement is most likely connected with the modification of the  surface properties.
%

The energy of the laser photon $\hbar\omega_L \approx 2.33$~eV is sufficiently below the typical value of the band gap of LLCCs, which varies from   4~eV to  2.8~eV for the typical chain lengths from 8 to 24 atoms available with the employed manufacturing facilities \cite{Mostaani}. It rules out a direct interband absorption as a possible source of excited carriers. However, the 532 nm laser light effectively excites localized plasmons in the Au NPs embedded in the carbyne film. These plasmons decay either radiatively by re-emitting absorbed photons or nonradiatively by transferring its energy to the electronic subsystem \cite{Brongersma2015,Clavero2013}. The interplay of these effects crucially depends on the material properties. In the nanoscale noble-metal structures, the dominating plasmon relaxation mechanism is the Landau dumping which leads to creation of high-energy carriers \cite{Khurgin2018}. This process excites electrons well above the Fermi level (with energies up to $\hbar \omega_L$ \cite{Clavero2013}), which is why plasmon-induced carriers are typically referred to as ``hot''. 

In the presence of continuous pumping, the hot-carrier generation rate is balanced by their relaxation via electron-hole recombination and thermal dissipation. The resulting non-equilibrium energy distribution of hot carriers $f_{\rm hot}(E)$ crucially depends on the excitation conditions and material properties \cite{Reddy2021,Manjavacas2014,Govorov2013} and can strongly deviate from the equilibrium one.  According to (\ref{Eq_LBcurrent}), the presence of hot carriers affects the tunneling current collected directly from the NP.
Note that the obtained CVCs are almost linear and symmetric in the dark, Fig.~\ref{fig1}f. It means that LDOSs of both tip and the NP are essentially constant within the bias swapping window while the tunneling matrix element $M(E)$ does not change appreciably. Therefore, the current-voltage dependence in the presence of laser light is governed by
\begin{equation}
I(U)\propto \int^{\infty}_{-\infty}\left[f_{\rm eq}(E) – f_{\rm hot}(E+eU) \right]dE.    
\end{equation}
Under the assumption of equilibrium distribution of hot carriers $f_{\rm hot}(E)=f_{\rm eq}(E,T_{\rm eff})$ at the effective temperature $T_{\rm eff}$ \cite{Reddy2021}, the current turns out to be independent of $T_{\rm eff}$ at any reasonable value of $k_{\rm B} T_{\rm eff}$ below Fermi energy. Indeed, $\int^{\infty}_{\infty} f_{\rm eq}(E) dE = k_{\rm B }T\ln{\left[1+\exp({E_{\rm F}/k_{\rm B}}T)\right]} \approx E_{\rm F}$. This clearly contradicts experimental results which exhibit a strong enhancement of tunneling under the light illumination.
Therefore, the observed phenomenon can be interpreted as a manifestation of a strongly non-equilibrium distribution $f_{\rm hot}(E)$ characterized by a significant asymmetry with respect to the Fermi level. This conclusion is in a good agreement with theoretically predicted energy distribution of the hot plasmon-induced carriers in metal NPs \cite{Govorov2013}.

A remarkable property of the plasmon-induced carriers is that they can leave NP and jump to the surrounding structure \cite{Clavero2013}. Various manifestations of this phenomenon include plasmon-assisted activation of chemical reactions with individual molecules on the NP surface \cite{Kazuma2018} and charge doping of two-dimensional structures  \cite{Fang2012,Kang2014}. 
The photoinduced injection of hot carriers is efficient provided that the electron energy distribution in the NP overlaps with the energy band or discrete energy levels of the neighbouring structure. In the case of NP-carbine bundle junction, it requires matching with the LUMO or HUMO states of the finite-length carbon chains attached to the NP surface, see Fig.~\ref{fig1}b. Note that the bundles are composed of the wires of different lengths which vary from a few unit cells to tens of unit cells. The positions of the HOMO and LUMO levels as well as the value of the energy gap strongly depend on the length of the straight parts of the chain. 
Neglecting the interaction between parallel chains, one can assume that the overlapped energy levels of the HOMO (LUMO) orbitals of individual chains form broad energy bands characterising the bundle as a whole. The overlap is amplified by the inhomogeneous broadening from the energy variation of the HOMO (LUMO) states characterizing individual chains distorted by the environment. 
Therefore, because of the presence of a finite energy gap, the band structure of the carbyne bundle is analogous to a bulk semiconductor. In this case, formation of the Schottky barrier on the NP/bundle interface is expected \cite{Clavero2013}. This barrier impedes the transfer of carriers with sub-barrier energy but is unable to stop injection completely as carriers can tunnel across the barrier. The charge separation at the NP/bundle interface is followed by their tunneling to the metallic substrate which is maintained at the ground potential during the STM study.
The charge regeneration in the metal NPs occurs effectively as they are in-touch with the substrate. However in the bundle, it is hindered by the Schottky barrier at the bundle/substrate interface. As a result, the excessive carriers are accumulated in the carbyne bundle leading to the effective doping of the a thin semiconductor film. 
Note that a similar electrical doping of the ultra-thin MoS$_2$ semiconductor and the graphene layer was reported in \cite{Fang2012,Kang2014}.

The transfer of carriers to the bundle modifies its properties and thereby affects the tunneling current flowing through the tip-bundle junction. The $I$--$U$ curves collected under light illumination demonstrate two characteristic features. The first one is a step-like increase of the current which occurs both at positive and at negative biases, see Fig.~\ref{fig1}d. According to Eq.~(\ref{Eq_LBcurrent}), the current flowing across the metal-semiconductor junction is crucially affected by the energy overlap between the LDOS of two electrodes \cite{ChenBook}. At zero bias, the Fermi level of the tip is aligned with the Fermi level of the bundle which is within the band gap. In this case $\rho_{\rm s}(E)$ vanishes and the weak current is mediated by the intra-gap defect states with low LDOS. However, as the tip Fermi level is biased above the conduction or below the valence band edges, the tunneling current increases sharply following the increase of the LDOS of the film. 

\begin{figure}
\includegraphics[width=0.9\linewidth]{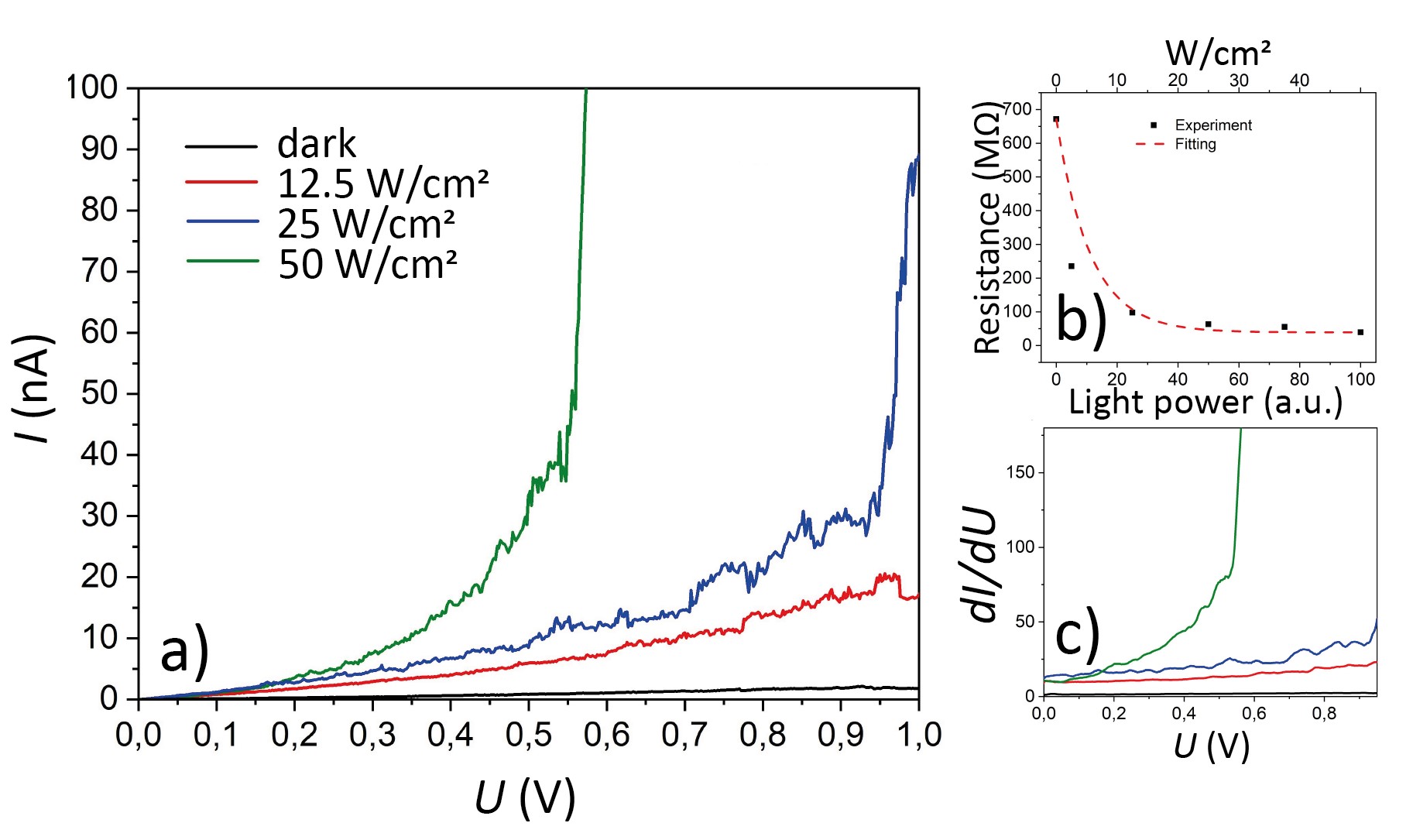}
\caption{ Evolution of the tunneling current through the NP-carbon thin film in the presence of a laser light. The main image shows current-voltage characteristics under various illumination intensities. The curves correspond to 25\%, 50\% and the total fluence of 50 $W/{cm}^2$. The top inset shows the differential form of the presented $I$--$U$  curves.  The bottom inset shows the experimental data (dots) and the exponential fit of the junction resistance at the near-zero bias. 
}\label{fig2}
\end{figure}

Therefore, position of the conductivity peaks provides an important information about the  band gap of the carbyne film \cite{Wiesendanger_book}. In this way, the $I(U)$-dependence shown in Fig.~\ref{fig1}d reveals a significant light-induced modification of the density of states in the film which corresponds to the effective narrowing of the band gap. The corresponding bias-dependence of the differential conductivity $dI/dU$ which quantifies LDOS \cite{ChenBook} is demonstrated in the upper inset on Fig.~\ref{fig2}. Note that a very similar effective reduction of the band gap has been observed under the laser light illumination of semiconductor quantum dots deposited on the Au substrate \cite{Sandhya2020}. This phenomenon can be connected with the presence of free carriers injected from the NP. In the low-dimensional semiconductors, the Coulomb screening provided by the external carriers has a strong effect on the band-structure properties \cite{Gao2017}. In particular, the strong reduction of the band gap by an amount of several hundreds of meV was reported in \cite{Qiu2019}.

Another important feature of the CVCs collected from the carbyne bundle film is their pronounced asymmetry. The sharp conductivity peaks corresponding to the valence and conduction band edges appear asymmetrically around zero bias. This behaviour is connected with the STM tip-induced band bending in the deposited film which is a characteristic feature of the STM studies of semiconductor surfaces  \cite{Feenstra2009}. The proximity of a metal tip triggers formation of an energy barrier of the height $\phi_b$ caused by the difference of the vacuum potentials of the tip and the sample \cite{Bell1988}. In this case, the position of the band edges are shifted by $\phi_b$ on the resulting $I(U)$-dependence. 

Although the barrier height depends on the tip shape and the local surface properties \cite{Quang2018}, it's magnitude at zero bias can be estimated as $\phi_b = A_{\rm tip} - \chi - (E_{\rm C} - E_{\rm Fs})$ is governed by the work function of the tip  $A_{\rm tip}$, the electron-to-vacuum affinity of the carbyne bundle $\chi$ and the gap between its Fermi level $E_{\rm Fs}$ and the bottom of the conduction band $E_{\rm C}$. Since injection of the plasmon-induced hot carriers affects the Fermi energy of the bundle $E_{\rm Fs}$, the band bending is enhanced under the light illumination resulting in a pronounced asymmetry of the CVC. The resulted $I(U)$-dependence collected under light illumination (see Fig.~\ref{fig1}d) corresponds to the $n$-type semiconductor-metal junction which is in agreement with the plasmon-induced electron injection hypothesis. 



The laser-intensity dependence of the tunnel current shown in Fig.~\ref{fig2} demonstrates a distinct correlation between the strength of excitation and the current amplitude. Such a correlation is anticipated assuming the localized plasmon field being a major source of carriers. The higher the field intensity, the quicker the rate of electron-hole pair creation due to the plasmon mode decay. The collected data demonstrate monotonic reduction of the junction resistance as the pump power increases. The experimental results can be fitted to the exponential dependence. We note, however, that the grows of the light intensity results not only in the increase of the magnitude of the electric current but the differential conductivity $dI/dU$ is affected as well. Figure \ref{fig2} demonstrates the growth of the differential conductivity with the increase of the light intensity.

\section*{Conclusion}

In conclusion, we demonstrated an efficient tool for a laser-induced generation of ``hot'' carriers in complex 2D systems consisting of bundles of linear carbon chains anchored with gold particles. Injection of hot electrons from the metal nanoparticles of a 10 nm diameter leads to the formation of an effective Schottky barrier at the junction between carbon chains and the conductive STM tip. In this regime, the value of the tunnel current  increases monotonically with increase of the excitation intensity. The observed phenomenon is a consequence of the transition of high-energy carriers from NPs to the carbyne wire where LUMO and LUMO levels of the carbon matrix strongly depend on the number of atoms in a single linear carbon chain.
These results demonstrate that the fabrication of monoatomic carbon wires end-capped with metal nanoparticles provides a versatile tool for harnessing of the fascinating properties of LLCCs. The fabrication of nanoribbons consisting of parallel monoatomic carbon chains is a promising step towards the realisation of nano-electronic circuits based on LLCC-gold complexes.

\section{Methods}
 We have used the method of laser fragmentation of colloidal carbon systems for creation of stabilised carbyne chains. The method is described in detail in \cite{Kutrovskaya20}. To create carbon-gold bonds we additionally illuminated the solution by nanosecond laser pulses generated by an Ytterbium (Yb) fiber laser with the central wavelength of 1.06~$\mu$m, the pulse duration of 100 ns, the repetition rate of 20 kHz and the pulse energy of up to 1 J. The sizes of gold NPs have been controlled by the dynamic laser scattering device Horiba SZ100.
The carbon phase was investigated with the use of Raman Senterra spectrometer. 
After the laser ablation stage, the solution containing LLCC fraction passed under the pressure of 70 kPa between two parallel plane electrodes biased with a voltage ranging from 100 to 1000 V and spaced by 1-2 cm. In the presence of a stationary electric field of the magnitude up to 10${}^{5}$ V/m, carbon chains became oriented along the field lines during the deposition on a solid substrate. To express the liquid phase removing the deposition was performed under IR heating at a constant temperature of 60 $^{\circ}$C $\pm$ 0.5 $^{\circ}$ C, where a control was provided by a laser pyrometer. During the deposition process, the nozzle was opened every 2 seconds for 0.3 s, the substrate was rotated at a frequency of 20 Hz, to ensure a deposited layer's uniformity.

For the detailed study of the LLCC-gold complexes, we have performed the high resolution transmission electron microscopy and X-ray diffraction studies using FEI Titan$^3$ with a spatial resolution of up to 2~{\AA}. Processing of TEM-images and diffraction patterns was conducted with the opened database package Image J 1.52 a. The tunneling measurements combined with atomic force microscopy have been performed using the Ntegra Aura nanolab device made by NT-MDT company.

\section{acknowledgments}
The work of S.K. and A.KA. is supported by the Westlake University (Project No. 041020100118) and by the Program 2018R01002 of Leading Innovative and Entrepreneur Team Introduction Program of Zhejiang and by Moscow Institute of Physics and Technology under the Priority 2030 Strategic Academic Leadership Program. I.C. acknowledges the support from the Grant no. MK-5318.2021.1.2. The work of A.P. is supported by the St. Petersburg State University (Project No. 94031307). This work was also partially supported by RFBR grants 17-32-50171, 18-32-20006 and 19-32-50095. The synthesis and deposition of LLCC have been performed at the Vladimir State University. Raman spectra was measured at the Center for Optical and Laser Materials Research, Research Park, St. Petersburg State University. TEM measurements have been performed in the ``System for microscopy and analysis'' LLC, Moscow. The authors thank M. Portnoi for valuable discussions of electrical properties of carbyne bundles.

\section* {Author Contributions}
\noindent S.K. has coordinated the collaborative work and analysed experimental data; 

\noindent I.C. contributed to the theoretical model, analysed the data, wrote and reviewed the manuscript;

\noindent A.O. performed the laser experiments;

\noindent V.S. realized the LLCC - Au complex deposition on a substrate;

\noindent A.L. contributed to the tunneling data collection;

\noindent A.P. performed the Raman spectrum collection and analyzed experimental data;

\noindent A.KA. contributed to the interpretation of the results.

\noindent A.K. has conceived the work, contributed to the synthesis of monoatomic carbon chains, realized HR-TEM microscopy study.


\end{document}